\documentclass[aps,prl,twocolumn,superscriptaddress]{revtex4}
\usepackage{graphicx}
\usepackage{dcolumn}
\usepackage{bm}

\begin{document}

\title{Electric field inversion asymmetry: Rashba and Stark effects for holes in resonant tunneling devices}

\author{H. B. de Carvalho}
\address{Grupo de Propriedades \'Opticas, Instituto de F\'{\i}sica
Gleb Wataghin, Universidade de Campinas, 13083-970, Campinas, SP,
Brazil.}
\author{M. J. S. P. Brasil}
\address{Grupo de Propriedades \'Opticas, Instituto de F\'{\i}sica
Gleb Wataghin, Universidade de Campinas, 13083-970, Campinas, SP,
Brazil.}
\author{V. Lopez-Richard}
\email{vlopez@df.ufscar.br}\address{Faculdade de Filosofia
Ci\^{e}ncias e Letras de Ribeir\~{a}o Preto, Departamento de
F\'{\i}sica e Matem\'{a}tica, Universidade de S\~{a}o
Paulo,14040-901, Ribeir\~{a}o Preto, SP Brazil.}
\author{I. Camps}
\address{Universidade Federal de S\~{a}o Carlos, Departamento de
F\'{\i}sica, 13560-905, S\~{a}o Carlos, SP, Brazil.}
\author{Y. Galv\~{a}o Gobato}
\altaffiliation{also at Departamento de F\'{\i}sica, Universidade
Federal de Santa Catarina.} \address{Universidade Federal de S\~{a}o
Carlos, Departamento de F\'{\i}sica, 13560-905, S\~{a}o Carlos, SP,
Brazil.}
\author{G. E. Marques}
\address{Universidade Federal de S\~{a}o Carlos, Departamento de
F\'{\i}sica, 13560-905, S\~{a}o Carlos, SP, Brazil.}
\author{L. C. O. Dacal}
\address{Instituto de Estudos Avan\c{c}ados, IEAv - CTA, C. P. 6044,
12231-970, S\~{a}o Jos\'{e} dos Campos, SP, Brazil.}
\author{M. Henini}
\address{School of Physics and Astronomy, University of
Nottingham, NG7 2RD, Nottingham, UK}
\author{L. Eaves}
\address{School of Physics and Astronomy, University of Nottingham,
NG7 2RD, Nottingham, UK}
\author{G. Hill}
\address{EPSRC National Centre for III-V Technologies,
University of Sheffield, Mappin Street, S1 3JD, Sheffield, UK}

\date{\today}

\begin{abstract}
We report experimental evidence of excitonic spin-splitting, in
addition to the conventional Zeeman effect, produced by a
combination of the Rashba spin-orbit interaction, Stark shift and
charge screening. The electric-field-induced modulation of the
spin-splitting are studied during the charging and discharging
processes of p-type GaAs/AlAs double barrier resonant tunneling
diodes (RTD) under applied bias and magnetic field. The abrupt
changes in the photoluminescence, with the applied bias, provide
information of the charge accumulation effects on the device.
\end{abstract}

\maketitle

The effect of the spin-orbit (SO) interaction in
quasi-two-dimensional (Q2D) systems has attracted renewed attention
in recent years. The topic has been on the focus of many optical and
transport investigations of spin-related phenomena in nanoscopic
systems~\cite{Mishenko,Tarasenko,Ganichev}, a subject of great
fundamental and technological
interest~\cite{Stepanenko,Kato,Rashba,Koga}. In this letter, we
address experimental evidence of electric field coupling to the spin
degree of freedom of carriers in RTD; here in particular, the
prevailing influence can be attributed to the SO and Stark effects
on the hole electronic structure. These interactions are relevant to
the study of the internal electric fields and the charge
accumulation in the structure. The simultaneous investigation of
optical and transport properties at high magnetic and electric
parallel fields, has permitted a thorough characterization of the
main processes involved in the system response. The novelty of this
result consists of the optical detection of electric field
modulation of the effective spin-splitting beyond the Zeeman effect
and its unambiguous correlation to the transport mechanisms which is
responsible for the charge buildup in the states of the RTD.

\begin{figure}[tbp]
\includegraphics[scale=0.47]{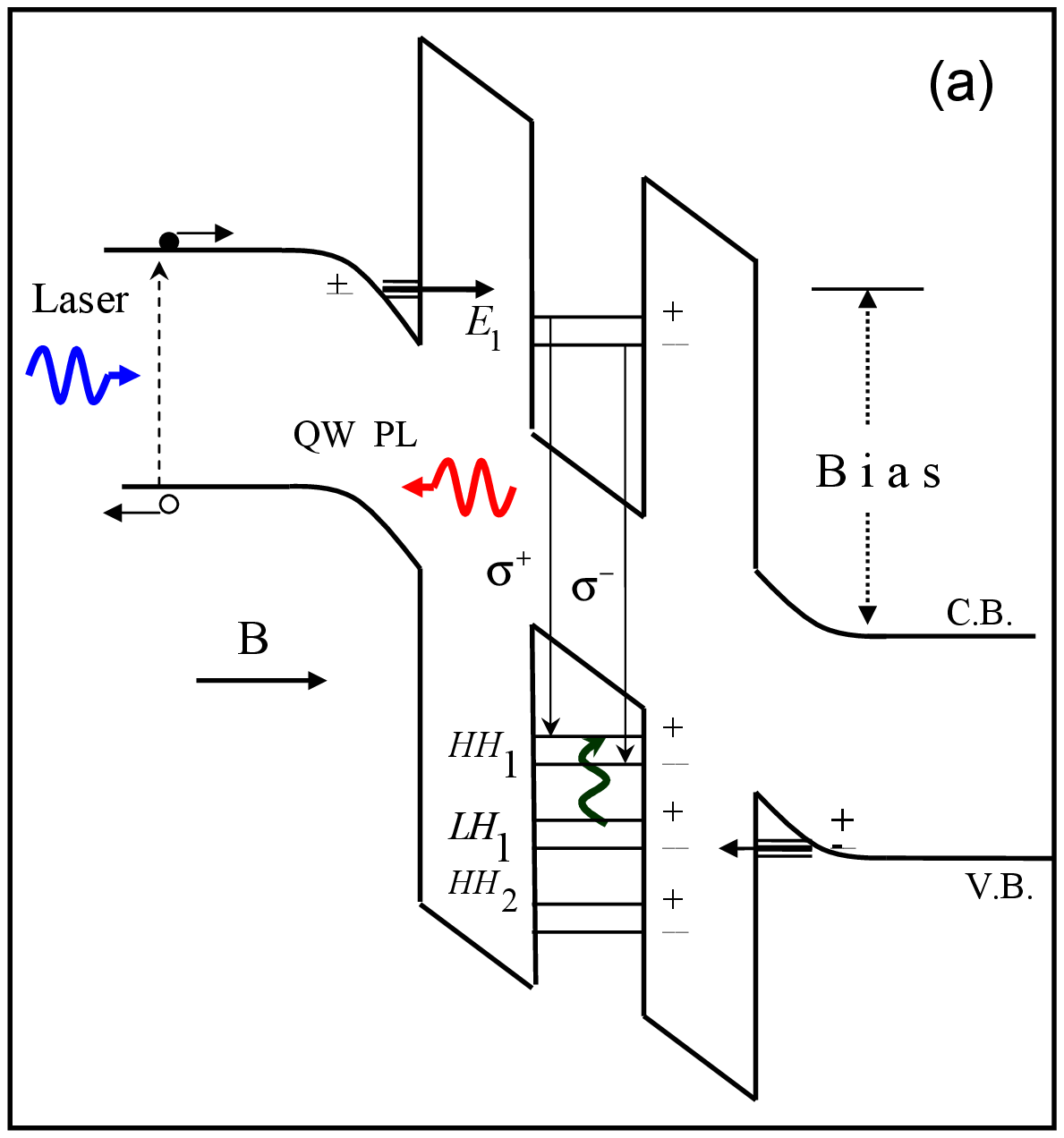}
\includegraphics[width=7.3 cm]{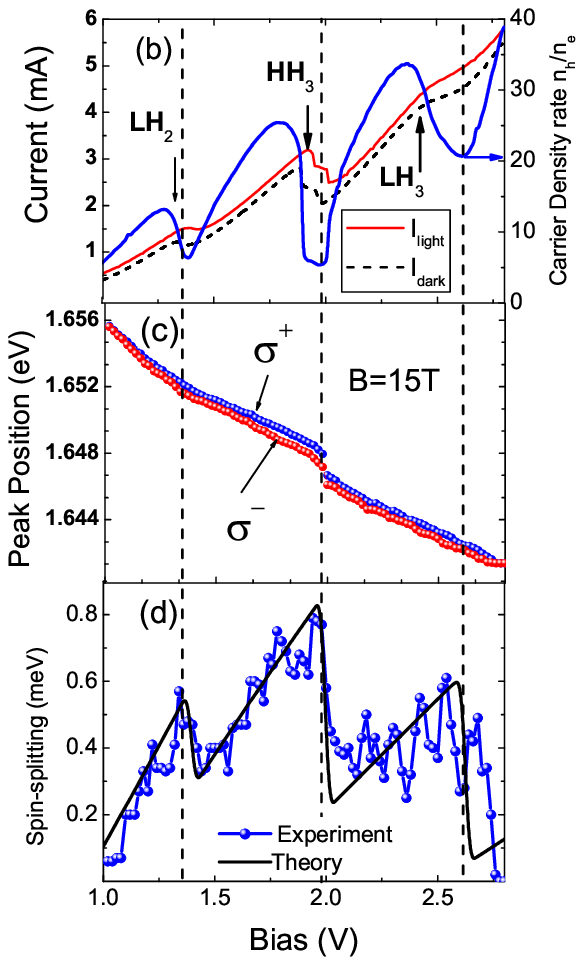}
\caption{Panel (a): Schematic diagrams for tunneling and
recombination processes in the $GaAs-AlAs$ RTD with 4.2nm (5.1nm)
well (barriers) width, under illumination. Below panels: Vertical
dashed-lines show the critical voltages where abrupt changes in the
properties occur for $B=15$ T. Panel (b): The $I-V$ characteristics
with ($I_{light}$) and without ($I_{dark}$) illumination. The
carrier density rate calculated through Eq.~(\ref{relation}). Panel
(c): Energy peak positions of $\sigma^{\pm}$ PL spectra as a
function of bias. Panel (d) Symbols: measured spin-splitting of
excitonic recombinations, for increasing bias. Solid-line:
Calculated spin-splitting generated by SO effects, in addition to
the conventional Zeeman effect.} \label{Fig1}
\end{figure}

This study is carried out on a symmetric $p-i-p$ $GaAs/AlAs$ RTD,
that has been previously used to characterize hole space charge
buildup and resonant effects in a magnetic field~\cite{Henini}. The
structure is in the form of a $400\mu m$ diameter mesa with a
metallic $AuGe$ annular top contact to allow optical access. The
diode was mounted in a superconducting magnet and the emission
spectra were recorded using a double spectrometer coupled to a CCD
system with polarizer facilities to select left (right)
$\sigma^{+(-)}$ configurations. When light from an $Ar^{+}$ laser is
focused close to the surface, minority electrons are
created~\cite{Henini}. As the bias approaches a resonant condition,
the carrier density inside the QW increases and then decreases,
resulting in the negative differential resistance (NDR) region when
the resonance is traversed. The photo-generated electrons tunneling
into the QW layer can recombine with the injected holes or tunnel
out of the well layer. These processes are represented schematically
in the Fig.~\ref{Fig1}~(a).

The $I-V$ characteristics, shown in Fig.~\ref{Fig1}~(b), displays a
series of peaks associated with the injected holes ($I_{dark}$) from
the hole accumulation layer formed in the outside interface of the
diode (see Fig.~\ref{Fig1}~(a)). Under illumination, an increase of
current is observed ($I_{light}$) due to the injection of minority
electrons. Under this condition, is noted the emergence of polarized
photoluminescence (PL) emissions from electron-hole (e-h) pair
recombinations inside the QW layer, as shown schematically in the
top panel. These PL spectra were recorded for $\sigma^{+}$ and
$\sigma^{-}$ polarizations as a function of the bias. The energy of
the PL peaks as well as the resulting spin-splitting as a function
of bias are shown in Figs.~\ref{Fig1}~(c) and (d), respectively.
We now focus on the mechanisms responsible for the abrupt changes
detected at the critical voltages, marked with dashed-lines, in
Fig.~\ref{Fig1}. As the bias (electric field) increases, both PL
polarized peak energies show the expected red-shift induced by the
quantum Stark effect, but also the sharp discontinuities at the
hole-resonant voltages. The abrupt variation of the PL-peak energies
has been attributed to changes in the QW level charge density near
the critical voltages; thus inducing changes on the built-in
electric field~\cite{Fisher}. These discontinuities are more evident
in the total spin-splitting energy vs bias, shown in
Fig.~\ref{Fig1}~(d).

Fig.~\ref{Fig1}~(d) shows fluctuations of the spin-splitting energy
and an increasing trend as bias is swept between two hole critical
voltages. The sharp decrease occurs when the NDR region is crossed
and the QW level is being discharged. Besides the usual Zeeman
effect, other two SO interactions modify the effective
spin-splitting: (i) the bulk inversion asymmetry (BIA or Dresselhaus
SO) induces spin-splitting only in zincblende heterostructures, (ii)
the structure inversion asymmetry (SIA or Rashba SO) causes
spin-splitting under applied electric field~\cite{Rashba1,gil}. Both
contributions depend on the material SO parameters and their effects
can be enhanced by the presence of a magnetic field~\cite{Rashba1}.
The properties addressed in this work occur at high electric field
(bias), then the Rashba SO contribution becomes dominant and we will
not consider the BIA SO term.

The complex behavior of our device is studied by modeling the states
by a square potential with a bias dependent electric field due to
the charge density built in the QW levels. The coupling between this
effective electric field and the spin degree of freedom is
introduced via the Rashba SO Hamiltonian for
electrons~\cite{Rashba1,gil} and for holes~\cite{Pala}. The kinetic
Luttinger Hamiltonian provides an accurate description of the
valence band admixture and permits to treat magnetic ($B$) and
electric ($F$) fields, as well as the Rashba SO in the same
framework. The full Hamiltonian, $H^{cond/val}=H^{L}\pm
I_{2j+1}\cdot eFz+H^{SO}$, may be decomposed into three terms: (i)
$H^{L}$ that describes the dynamics associated with the Landau and
Zeeman quantization in a magnetic field, (ii) $I_{2j+1}\cdot eFz$
contains all Stark shift-related terms that produce the inversion
asymmetry induced by the electric field ($F$); here $I_{2j+1}$ is
the $(2j+1)$-rank unity matrix, and (iii) the Rashba SO term that
couples the dynamical linear momentum with the spin degree of
freedom. The SO term is treated in the $(2j+1)$-rank representation
of the total momentum with $j=1/2$ for electrons and $j=3/2$ for
holes~\cite{Pala} as
\begin{equation}
H_{\gamma}^{SO}=\alpha_{\gamma}\frac{\sqrt{2}F}{\lambda_{c}}i\left(
aJ_{+}-a^{\dag}J_{-}\right).
\end{equation}
Here $\alpha_{cond(val)}$ is the Rashba SO parameter for conduction
(valence) band, $\lambda_{c}$ is the magnetic cyclotron radius, and
$J_{\pm}=\frac{1}{2}\left(J_{x}\pm J_{y}\right)$, where $J_{i}$ is
the $4\times4$ ($2\times2$) angular momentum matrix for holes
(electrons). One advantage of this SO representation is that it
permits a wavefunction expansion in each component of the spinor
state with the required symmetries defined by the \textit{lateral}
and \textit{vertical} confinement types. The basis set also combines
the band edge periodic Bloch functions in the total momentum
representation: $\left\vert s\uparrow\downarrow\right\rangle$,
$\left\vert hh\uparrow\downarrow\right\rangle$, $\left\vert
lh\uparrow\downarrow\right\rangle$, the vertical eigenstates:
$A_{2k-1}(z)$-even and $A_{2k}(z)$-odd parity for $k=1,2,...$; and
the lateral Landau states $\left\vert N\right\rangle$. The
eigenfunctions for the conduction ($\Phi_{c}$) and for the valence
band states ($\Phi_{v}$) have the general form,
\begin{equation}
\Phi _{c}=\left[
\begin{array}{c}
A_{1}^{c}\left\vert N\right\rangle \left\vert s\uparrow
\right\rangle
\\
A_{1}^{c}\left\vert N+1\right\rangle \left\vert s\downarrow
\right\rangle  \\
A_{2}^{c}\left\vert N\right\rangle \left\vert s\uparrow
\right\rangle
\\
A_{2}^{c}\left\vert N+1\right\rangle \left\vert s\downarrow
\right\rangle  \\
A_{3}^{c}...%
\end{array}%
\right],%
\Phi _{v}=\left[
\begin{array}{c}
A_{1}^{v}\left\vert N-2\right\rangle \left\vert hh\uparrow
\right\rangle  \\
A_{1}^{v}\left\vert N-1\right\rangle \left\vert lh\uparrow
\right\rangle  \\
A_{1}^{v}\left\vert N\right\rangle \left\vert lh\downarrow
\right\rangle  \\
A_{1}^{v}\left\vert N+1\right\rangle \left\vert hh\downarrow
\right\rangle  \\
A_{2}^{v}\left\vert N-2\right\rangle \left\vert hh\uparrow
\right\rangle  \\
A_{2}^{v}\left\vert N-1\right\rangle \left\vert lh\uparrow
\right\rangle  \\
A_{2}^{v}\left\vert N\right\rangle \left\vert lh\downarrow
\right\rangle  \\
A_{2}^{v}\left\vert N+1\right\rangle \left\vert lh\downarrow
\right\rangle  \\
A_{3}^{v}...%
\end{array}%
\right].\label{waves}
\end{equation}
Note that $N=-1,0,1,2,...$ is an effective Landau index labeling the
\textit{lateral} confinement symmetries (parities) of each
component. The sequence of periodic valence Bloch states in the
components is determined by the sequence chosen to write the
Luttinger Hamiltonian. These vector states have, in principle,
infinite dimension since the index $k$, used to enumerate the
\textit{vertical} functions $A_{2k-1}$ and $A_{2k}$, runs over all
positive integer numbers.

\begin{figure}[tbp]
\includegraphics[width=8. cm]{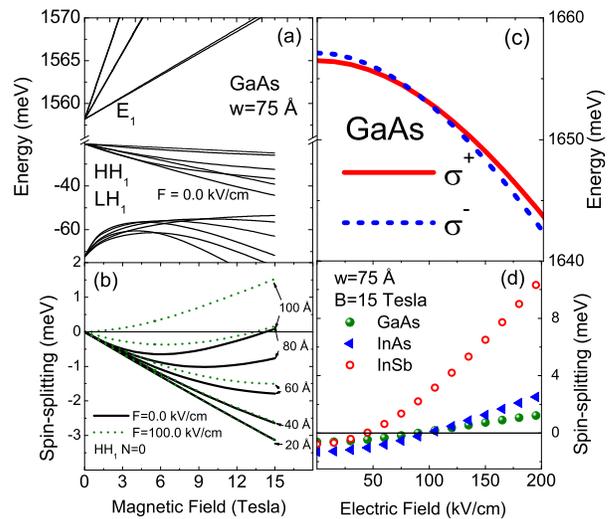}
\caption{Panel (a): Landau level fan charts ($N=0,1,2$) for the
first electron ($E_1$) and hole ($HH_1$, $LH_1$) subbands in a
$GaAs$ QW with width $w=75 \mathring{A}$ and $F=0$. Panel (b):
Calculated Zeeman splitting of the $HH_{1}$ ground state. The
solid-lines (dashed-lines) indicate the results without (with) the
inversion asymmetry induced by the electric field. Panel (c):
Interband transition energies for $\sigma^{\pm}$ polarized light
emissions in a $GaAs$ QW, with $B=15$ Tesla. Panel (d): The
additional SO contribution to the calculated spin-splitting energy
of excitonic recombination in identical QW's of different III-V
compounds, as a function of the electric field.} \label{Fig2}
\end{figure}

Fig.~\ref{Fig2}~(a) shows the calculated electron and hole magnetic
dispersions for the lowest energy subband of carriers in the
conduction ($E_1$) and in the valence($HH_1$, $LH_1$) bands of the
$GaAs$ QW. Note that the effective Zeeman splitting for holes and
electrons can be tuned by the external fields $F$ and $B$. This also
reflects the strong admixture of states generated by the combination
of Rashba SO plus Stark effects, in particular, for the valence band
where the strong nonparabolicity is present in the kinetic Luttinger
Hamiltonian. A comparison between the contribution from the
nonparabolicity ($F=0$) and from SO plus Stark effects ($F=100$
kV/cm) to the effective spin-splitting of the $HH_1$ ground state,
is shown in Fig.~\ref{Fig2}~(b) for different degrees of
\textit{vertical} confinement. For fixed $B$-value, an increase of
well width increases the energy splitting since the coupling between
states with close energy is enhanced. The corresponding interband
transition energies, calculated for a $GaAs$ QW, are displayed in
Fig.~\ref{Fig2}~(c) as a function of the electric field. The
inversion asymmetry, introduced by the electric field, effectively
couples QW subbands with different \textit{lateral} and
\textit{vertical} symmetries. This coupling leads to the
spin-splitting modulation by the electric field, which is the main
topic addressed. Note that the strength of modulation can be
enhanced for materials with larger SO parameters as shown in
Fig.~\ref{Fig2}~(d). The larger is the Rashba SO parameter, such as
in $InAs$ or $InSb$, the stronger is the coupling between the
electric field and the spin-degree of freedom.

Superimposed to the external electric field (bias), there is an
internal field component due to the charge distribution throughout
the diode structure. The relation between the applied bias and the
total effective electric field $F$ experienced by the carriers would
require a selfconsistent calculation. Here we consider an average
effective field which qualitatively accounts for the main effects of
the inversion asymmetry on the spin-splitting energy seen in the PL
emissions. From $I-V$ characteristics with and without illumination
[see Fig.~\ref{Fig1}~(a)] we may estimate the ratio between the
majority ($n_{h}$) and minority ($n_{e}$) carrier densities inside
the QW, as
\begin{equation}
\frac{n_{h}}{n_{e}}=2
\frac{I_{dark}\left(V\right)}{I_{light}\left(V\right)-I_{dark}\left(V\right)}.
\label{relation}
\end{equation}
Together with the $I-V$ characteristics, this ratio is shown in
panel (b) of Fig.~1, as a function of bias and for the highest laser
excitation power. Note that in our experimental conditions the
density of holes (majority) is always larger than the density of
electrons (minority). Moreover, $n_{h}/n_{e}$ increases smoothly
during the charge buildup process, as the bias is being swept
between two hole resonant voltages and an abrupt drop occurs during
the QW discharge (NDR region).

Owing to the smooth variation of $n_{h}/n_{e}$ and to the
experimental condition $n_{h}\gg n_{e}$, we have chosen to use a
total field $F$ changing linearly with an effective charge density,
$n_{h}^{eff}$, and giving rise to an uniform field. In the presence
of an asymmetry induced by the external field, this is a fair
approximation~\cite{Stern,Winkler,Landi}, and can be written
\begin{equation}
F=F^{ext}+\frac{e}{\epsilon}n_{h}^{eff}, \label{electric}
\end{equation}
where $\epsilon=13.18$ is the static dielectric constant for $GaAs$.
All the other parameters for $GaAs$ used in the simulation, such as
effective masses, Luttinger and Rashba parameters were taken from
the literature.

Based on these assumptions we have estimated the asymmetry effects
on the modulation of spin-splitting seen in the PL emissions, due to
the charge accumulation in the QW levels. This is compared with
experimental results in Fig.~\ref{Fig1}~(d). A percentage of the
experimental wiggly increase is due to charge fluctuations on
system. Our theoretical model does not take into account any type of
charge fluctuations.

For this Q2D hole-rich gas, the PL emissions must arise from the
recombination of either positive trions ($X^{+}$) or neutral
excitons ($X^{0}$)~\cite{Yara,Eaves}. In order to understand the
behavior of the PL peaks in Fig.~\ref{Fig1}~(c) with abrupt
variations at resonant transitions for both spin-up and spin-down
recombinations, we must consider the charge density-dependent
screening on the binding energy of the excitonic complexes involved
in the PL emissions~\cite{Dacal}. An increase in the Q2D carrier
density leads to screening of the Coulomb interaction that
inevitably weakens the binding energy of the excitonic
complexes~\cite{Dacal}. Moreover, the charge buildup also gives rise
to modulations to: (i) the e-h Coulomb interaction and; (ii) the
Stark shift. Cases (i) and (ii) give opposite contributions to the
optical transition energies. Near the critical voltages, case (ii)
produces a blue-shift due to the reduction of the effective electric
field, whereas case (i) induces a red-shift due to the decrease of
the excitonic binding energies. These competing effects give origin
to the abrupt changes in the peak position of $\sigma^+$ and
$\sigma^-$ PL emissions of Fig.~\ref{Fig1}~(c). Therefore, the
spin-splitting energy is not affected by these excitonic
corrections.

As final comments, the good agreement with experimental increasing
trend required an effective density in the range $6\cdot10^{12}
cm^{-2}<n_{h}^{eff}<14\cdot 10^{12} cm^{-2}$. This is larger than
the values reported in Ref.~\cite{Henini} using a different relation
between bias and local fields. The larger value for the effective
hole charge density is due to the neglected contribution coming from
the hole density in interface accumulation layer. Furthermore, the
effective QW width $w=75 \mathring{A}$ was adjusted to fit the $e-h$
pair recombination energy at $F=0$. In this $p-i-p$ sample, the
modulation of the excitonic spin-splitting is strongly related to
valence band admixture. As noted in Fig.~\ref{Fig2}~(a), the set of
$HH_1$ and $LH_1$ Landau dispersions are highly nonparabolic, in
contrast to the almost parabolic conduction band levels. This effect
is strongly affected by the QW width that determines the separation
between the coupled $HH_1$ and $LH_1$ subbands, as shown in
Fig.~\ref{Fig2}~(b).

In summary, we have observed that spin-splitting dependence on bias
is mainly influenced by the renormalization of the hole charge
density in the uniform field approximation. This spin-splitting
induced by the combined effects of Rashba SO interaction, Stark
effect, Zeeman and interband couplings acts as a probe of the
relation between the fluctuating internal field $F$ and the
continuously varying bias $V$. We have shown how the electronic
structure is affected by the modification of the effective field $F$
in our RTD and by the modulation of the Rashba SO and screening
effects induced by charge fluctuations. We have studied these
effects in simultaneous optical and transport measurements of the
main physical processes involved.

The authors acknowledge financial support from Brazilian agencies
FAPESP and CNPq and from the UK Engineering and Physical Sciences
Research Council.


\begin{thebibliography}{99}
\bibitem{Mishenko} E. G. Mishchenko, A. V. Shytov, B. I. Halperin,
Phys. Rev. Lett. \textbf{93}, 226602 (2004).

\bibitem{Tarasenko} S.A. Tarasenko, V.I. Perel', I.N. Yassievich, Phys. Rev.
Lett. \textbf{93}, 056601 (2004).

\bibitem{Ganichev} S.D. Ganichev, V.V. Bel'kov, L.E. Golub, E.L. Ivchenko,
P. Schneider, S. Giglberger, J. Eroms, J. De Boeck, G. Borghs, W.
Wegscheider, D. Weiss, W. Prettl, Phys. Rev. Lett. \textbf{92},
256601 (2004).

\bibitem{Stepanenko} D. Stepanenko, N. E. Bonesteel, Phys. Rev. Lett. \textbf{93}, 140501 (2004).

\bibitem{Kato} Y. Kato, R. C. Myers, A. C. Gossard, D. D. Awschalom, NATURE
\textbf{427}, 50 (2004).

\bibitem{Rashba} E.I. Rashba, A.L. Efros, Phys. Rev. Lett. \textbf{91}, 126405
(2003).

\bibitem{Koga} T. Koga, J. Nitta, H. Takayanagi, Phys. Rev. Lett.
\textbf{88}, 126601 (2002).

\bibitem{Henini} R.K. Hayden, L. Eaves, M. Henini, D.K. Maude and J.C.
Portal, G. Hill, Appl. Phys. Lett. \textbf{60}, 1474 (1992).

\bibitem{Fisher} T.A. Fisher, P.D. Buckle, P.E. Simmonds, R. J.
Teissier, S. Skolnick, C.R.H. White, D.M. Whittaker, L. Eaves, B.
Usher, P.C. Kemeny, R. Grey, G. Hill, and M.A. Pate, Phys. Rev. B
\textbf{50}, 18469 (1994).

\bibitem{Rashba1} Y.A. Bychkov, E.I. Rashba, J. Phys. C: Solid State Physics \textbf{17}, 6039
(1984).

\bibitem{gil} G.E. Marques, A.C. Bittencourt, C.F. Destefani, and S.E. Ulloa,
Phys. Rev. B \textbf{72}, 045313, (2005).

\bibitem{Pala} Marco G. Pala, Michele Governale,
J\"{u}rgen K\"{o}nig, Ulrich Z\"{u}licke,
and Giuseppe Iannaccone, Phys. Rev. B \textbf{69}, 045304 (2004).

\bibitem{Stern}F. Stern, S. Das Sarma, Phys. Rev. B \textbf{30} 840 (1984).

\bibitem{Winkler} R. Winkler, H. Nohb, E. Tutucb, and M. Shayeganb; Physica E \textbf{12},
428 (2002).

\bibitem{Landi} S.M. Landi, C.V.-B. Tribuzy, P.L. Souza, R. Butendeich, A.C. Bittencourt, and G.E.
Marques, Phys. Rev. B \textbf{67}, 085304 (2003).

\bibitem{Yara} A. Vercik, Y. Galv\~{a}o Gobato, I. Camps, G.E. Marques, M.J.S.P.
Brasil, and S.S. Makler, Phys. Rev. B \textbf{71}, 075310 (2005).

\bibitem{Eaves} F.J. Teran, L. Eaves, L. Mansouri, H. Buhmann, D.K. Maude,
M. Potemski, M. Henini, and G. Hill Phys. Rev. B \textbf{71},
161309R (2005).

\bibitem{Dacal} L.C.O. Dacal and J.A. Brum, Phys. Rev. B \textbf{65},
115324 (2002). Also, J.A. Brum, G. Bastard, and C. Guillemot, Phys.
Rev. B \textbf{30}, 905 (1984). L.C.O. Dacal and J.A. Brum, Physica
E \textbf{12} 546 (2002).

\end{thebibliography}
\end{document}